\documentclass[a4paper]{jpconf}
\usepackage{graphicx}
\begin{document}

\title{Supernovae constraints on dark energy and modified gravity models}

\author{M. C. Bento, O. Bertolami, 
N. M. C. Santos \footnote[1]{Present address: Institut f\"ur
Theoretische Physik, Universit\"at
Heidelberg, Philosophenweg 16, 69120 Heidelberg, Germany} and
A. A. Sen \footnote[2]{Present address: Department of Physics and Astronomy, 
Vanderbilt University, Nashville TN37235, USA}}

\address{ Departamento de
F\'{\i}sica, Avenida Rovisco
Pais, 1049-001 Lisboa, Portugal}

\ead{bento@sirius.ist.utl.pt, orfeu@cosmos.ist.utl.pt,
 n.santos@thphys.uni-heidelberg.de, anjan.sen@vanderbilt.edu}

\begin{abstract}
We use the Type Ia Supernova {\it gold} sample to constrain the
 parameters of dark energy models namelly the Cardassian, Dvali-Turner (DT)
 and generalized Chaplygin gas (GCG) models. In our best fit analysis for
 these dark energy proposals we consider flat and the non-flat
 priors. For all models, we find that relaxing the flatness condition
 implies that data favors a positive curvature; moreover, the
 GCG model is nearly flat, as required by Cosmic
 Microwave Background (CMB) observations.

\end{abstract}

\section{Introduction}

Various proposals have been put forward to explain recent
observations indicating that the universe is accelerating.
A possible explanation is that the universe is filled with dark energy
in the form of an exotic component, the 
generalized Chaplygin gas, with negative equation of state
\cite{Kamenshchik:2001cp,Bilic:2001cg,Bento:2002ps}. The striking
feature of this model is that it allows for an unification of dark
energy and dark matter \cite{Bento:2002ps}.

Another possible explanation for the accelerated expansion of the
Universe could be the infrared modification of  gravity one should
expect from  extra dimensional physics, which would
lead to a modification of the effective Friedmann equation at late
times. A concrete model has been suggested by Dvali, Gabadadze and
Porrati  \cite{Dvali} and later generalized by Dvali and Turner
\cite{Dvali:2003rk}. Another
possibility is the modification of the Friedmann equation by the
introduction of an additional nonlinear term proportional to
$\rho^n$, the so-called Cardassian model \cite{Freese:2002sq}.

Currently type Ia supernovae (SNe Ia) observations provide the
most direct way to probe the dark energy component at low to
medium redshifts.  Recently,
supernovae data has been analysed by various groups
 and it was shown that it yields relevant
constraints on some cosmological parameters. In particular, it is
possible to conclude that, when one considers the full supernova
data set, the decelerating model is ruled out with a significant
confidence level \cite{Choudhury:2003tj}.
 It is also shown that one can measure the
current value of the dark energy equation of state with higher
accuracy and the data prefers the phantom kind of equation of
state, $w_X < -1$. Furthermore, the most significant result of
that analysis is that, without a flat prior,  supernovae data
does not favour a flat $\Lambda$CDM model at least up to
$68\%$ confidence level,
which is consistent with other cosmological observations. In what concerns
the equation of state of the dark energy component,  it has been
shown in Ref. \cite{Alam:2003}, using the same set of
supernovae data, that  the best fit equation of state of dark energy evolves
rapidly from $w_X \simeq 0$ in the past to $w_X \sim -1$ at
present, which suggests that a time varying dark energy fits the data better
 than the  $\Lambda$CDM model.

In this paper, we analyze the Cardassian, the DT and the GCG
models using so called {\it gold} sample SNe Ia compilation of
data by  Riess {\it et al.} \cite{Riess:2004nr}; we
 consider both flat and non-flat
priors. For more details see Ref. \cite{Bento:2004ym}.

\section{Observational constraints from supernovae data}

The observations of supernovae measure essentially the
apparent magnitude $m$, which is related to the (dimensionless) 
luminosity distance $D_L$ by
\begin{equation}
m(z) = {\cal M} + 5 \log_{10} D_L(z) ~,
\end{equation}
where
\begin{equation}
D_L(z) \equiv {H_0}(1 + z) \int_0^z {{1}\over{H(z')}} dz'~.
 \label{DL}
\end{equation}
Also,
\begin{equation}
{\cal M} = M + 5 \log_{10}
\left({{c/H_0}\over{1~\mbox{Mpc}}}\right) + 25~,
\end{equation}
where $M$ is the absolute magnitude which is believed to be constant for
all supernovae of type Ia.
For our analysis, we consider the set of supernovae data
recently compiled  by Riess {\it et al.} \cite{Riess:2004nr} known as
the {\it gold} sample.
The data points in this sample are  given in terms of the distance modulus
$\mu_{\rm obs}(z) \equiv m(z) - M_{\rm obs}(z)$
and the $\chi^2$ is calculated from
\begin{equation}
\chi^2 = \sum_{i=1}^n \left[ {{\mu_{\rm obs}(z_i) - {\cal M}' - 5
\log_{10}D_{L \rm th}(z_i; c_{\alpha})}\over{\sigma_{\mu_{\rm
obs}}(z_i)}} \right]^2~,
 \label{chisq2}
\end{equation}
where ${\cal M}' = {\cal M} - M_{\rm obs} $ is a free parameter
and $D_{L \rm th}(z; c_{\alpha})$ is the theoretical prediction
for the dimensionless luminosity distance of a supernova at a
particular distance, for a given model with parameters
$c_{\alpha}$. The errors $\sigma_{\mu_{\rm obs}}(z)$  take into
 account the effects of peculiar motions.
Minimization of the $\chi^2$, with respect to
${\cal M}'$, $\Omega_{m}$, $\Omega_k$  and the respective model
parameter(s) leads to the results  summarized in Table \ref{table:best}. 
The best fit value for ${\cal M}'$ is 43.3 for all  models.


\begin{table}[t!]
\caption{\label{table:best}Best fit parameters for the Cardassian, DT and GCG
models, considering flat and non-flat priors.}
\begin{center}
\begin{tabular}{l l l l l}
 \hline \hline
 \hline
 {\bf Cardassian model} & $\Omega_m$  & $n$ & $\Omega_k$ &
  $\chi^2$ \\
\hline
Flat  Prior  & $0.49$ & $-1.4$ & $-$ & $173.7$ \\
Non-Flat Prior  & $0.21$ & $-3.1$ & $0.47$ & $173.2$ \\
\hline
 {\bf  DT model}  & $\Omega_m$  &$\beta$ &$\Omega_k$ & $\chi^2$ \\
 \hline
Flat  Prior  & $0.51$ & $-19.2$ & $-$ & $174.7$ \\
Non-Flat Prior  & $0.24$ & $-60.0$ & $0.43$ & $174.0$ \\
\hline
 {\bf GCG model}  & $A_s$ & $\alpha$&$\Omega_k$ &  $\chi^2$\\
 \hline
Flat  Prior   & $0.93$ & $2.8$ & $-$ & $174.2$ \\
Non-Flat Prior  & $0.97$ & $4.0$ & $0.02$ & $174.5$ \\
 \hline\hline
\end{tabular}
\end{center}
\end{table}

\section{Cardassian model}

We first consider the  Cardassian model \cite{Freese:2002sq};
in this model, the universe is composed only of radiation and matter
 and the increasing expansion
rate is given by
\begin{equation}
\label{FriedCmodel} H^2= {{8 \pi}\over {3 M_{\rm Pl}^2}} \left(
\rho + b \rho^n \right) - {{k} \over {a^2}}
\end{equation}
where $M_{\rm Pl} = 1.22 \times 10^{19}$GeV is the $4$-dimensional
Planck mass, $b$ and $n$ are constants, and we have added a
curvature term to the original Cardassian model. The new term dominates only
recently, hence in order to get the  recent acceleration
in the expansion rate, $n < 2/3$ is required.
\begin{figure}[ht]
\includegraphics[width=18pc]{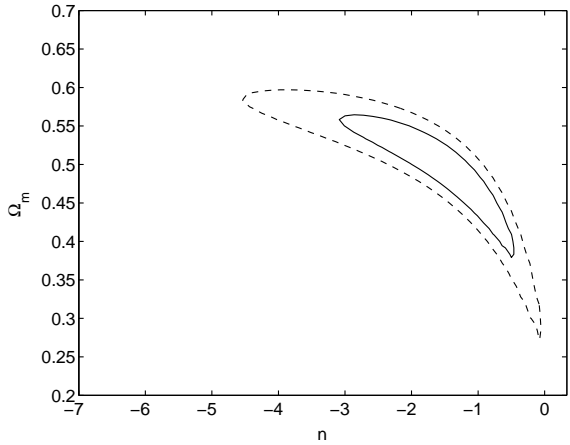}\hspace{2pc}%
\begin{minipage}[b]{16pc}
\caption{\label{fig:flatcard} Confidence
    contours  for the flat
  Cardassian model. The solid and dashed lines represent the $68\%$
  and $95\%$ confidence regions, respectively.}
\end{minipage}
\end{figure}


 At present, the
universe is matter dominated and  Eq. (\ref{FriedCmodel})
can be rewritten as
\begin{equation}
\label{FriedCmodel2} \left({{H} \over {H_0}}\right)^2= \Omega_{m}
(1+z)^3+\Omega_k (1+z)^2 \nonumber \\
+ (1-\Omega_{m}-\Omega_k)(1+z)^{3n}~,
\end{equation}
where $H_0$ is the present value of the Hubble constant and
$\Omega_k=-{{k} \over {a_0^2 H_0^2}}$ is the present curvature
parameter. Notice that  the case $n=0$ corresponds to the $\Lambda$CDM
model.
\begin{figure}[h]
\begin{minipage}{37pc}
 \includegraphics[width=18pc]{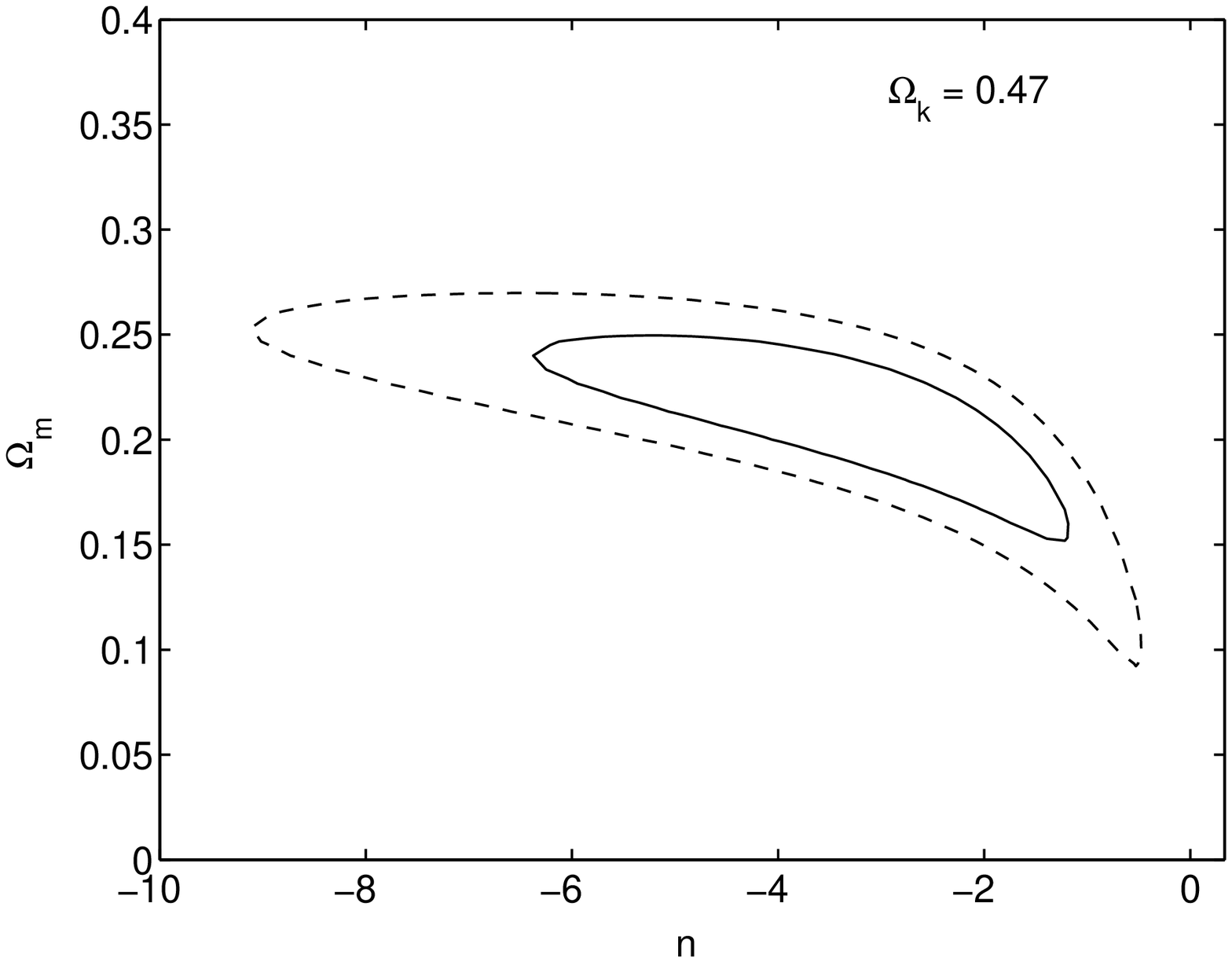}
\includegraphics[width=18pc]{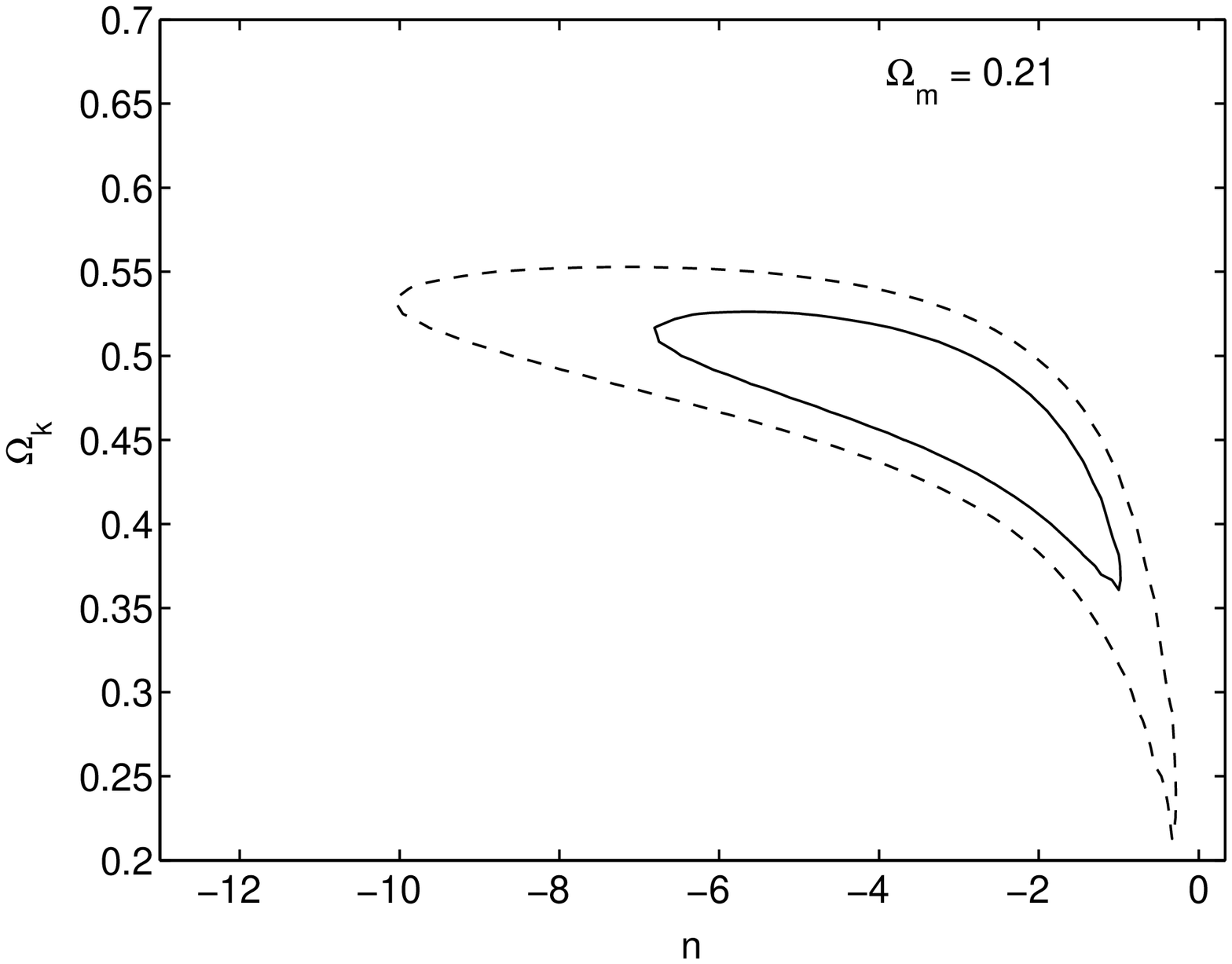}
\caption{\label{fig:card}  As for Figure \ref{fig:flatcard}, for the
  non-flat Cardassian model.}
\end{minipage}
\end{figure}

For the flat case, we have only two parameters and their best
fit values  are $\{\Omega_{m},n\}=\{0.49,-1.4\}$.
 In Fig. \ref{fig:flatcard} we show the 68$\%$ and 95$\%$
confidence contour plots, where it is clear that the $\Lambda$CDM model
 is excluded at a 95$\%$
confidence level.

If we relax the flat prior, we find that the
 best fit values are $\Omega_k = 0.47$,
$\Omega_m = 0.21$ and $n=-3.1$. Moreover (see Fig. \ref{fig:card})
 the  case $n=0$ is again excluded at $95\%$ confidence level.

\begin{figure}[ht]
 \includegraphics[width=18pc]{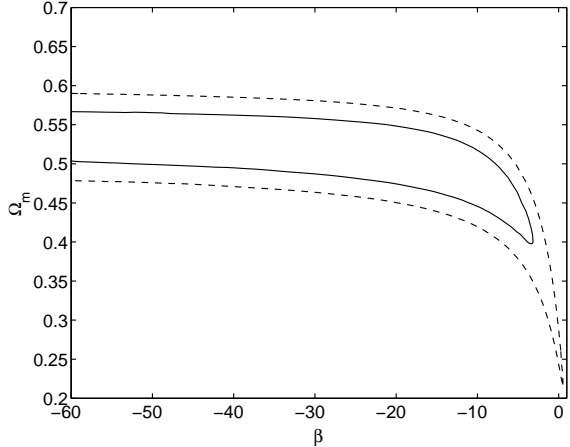}\hspace{2pc}%
\begin{minipage}[b]{16pc}
  \caption{\label{fig:flatDT} Confidence contours for
    the flat DT model. The solid and dashed lines represent the $68\%$
  and $95\%$ confidence regions, respectively.}
\end{minipage}
\end{figure}


\section{Dvali-Turner model}

The second model we consider is the one proposed by Dvali \&
Turner \cite{Dvali:2003rk}, where the Friedmann equation is
modified by the addition of a new term which  may
 arise in theories with extra dimensions
\cite{Dvali}

\begin{equation}
H^2 \, -  \, {H^{\beta} \over r_c^{2 - \beta}} \, = \, {8\pi \over
3 M_{\rm Pl}^2} \rho - {{k} \over {a^2}}~,
\end{equation}
where $r_c$ is a crossover scale which sets the
scale beyond which the laws of the 4-dimensional gravity breakdown
and become 5-dimensional  and we have  introduced a curvature
term. 
\begin{figure}[h]
\begin{minipage}{37pc}
 \includegraphics[width=18pc]{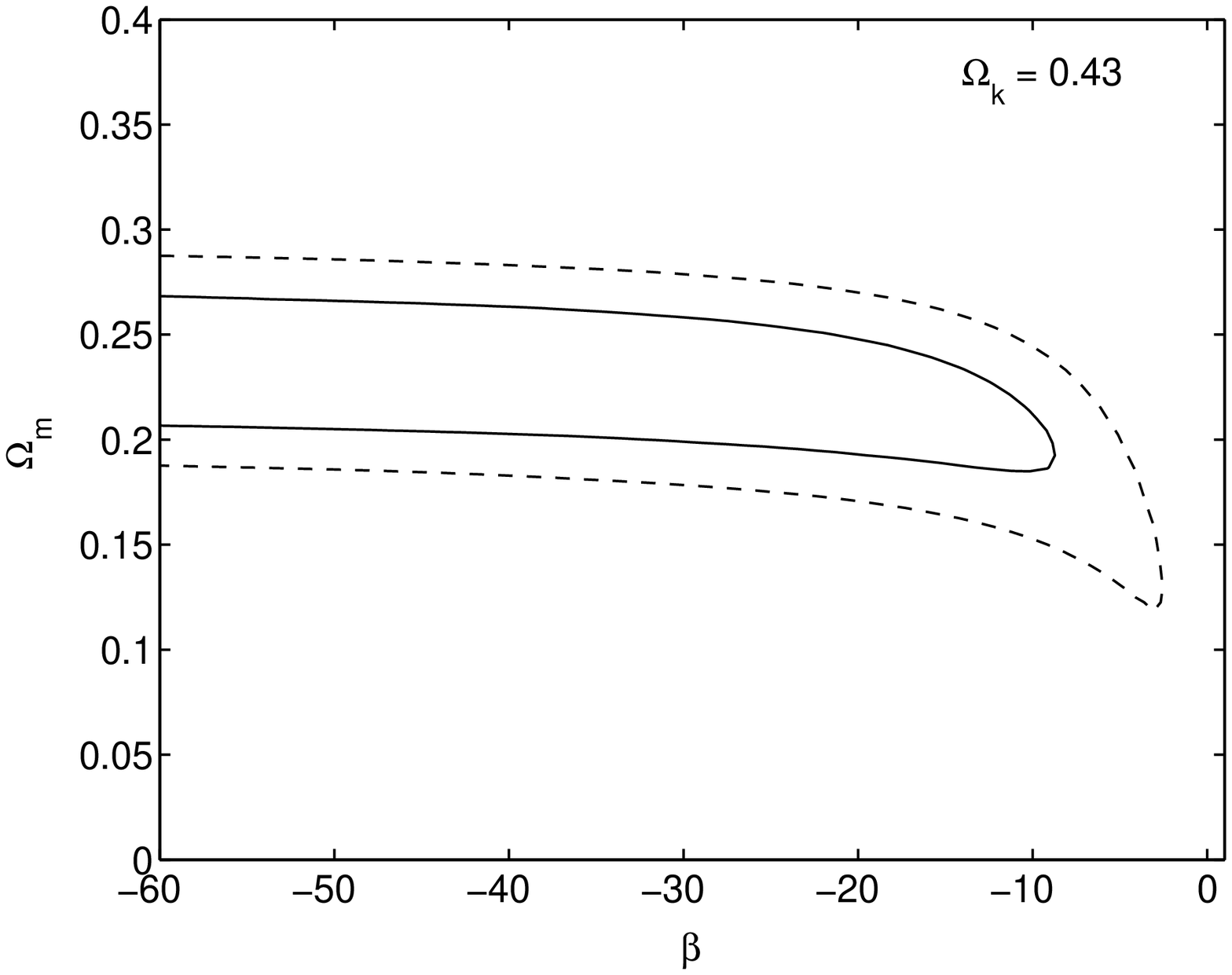}
 \includegraphics[width=18pc]{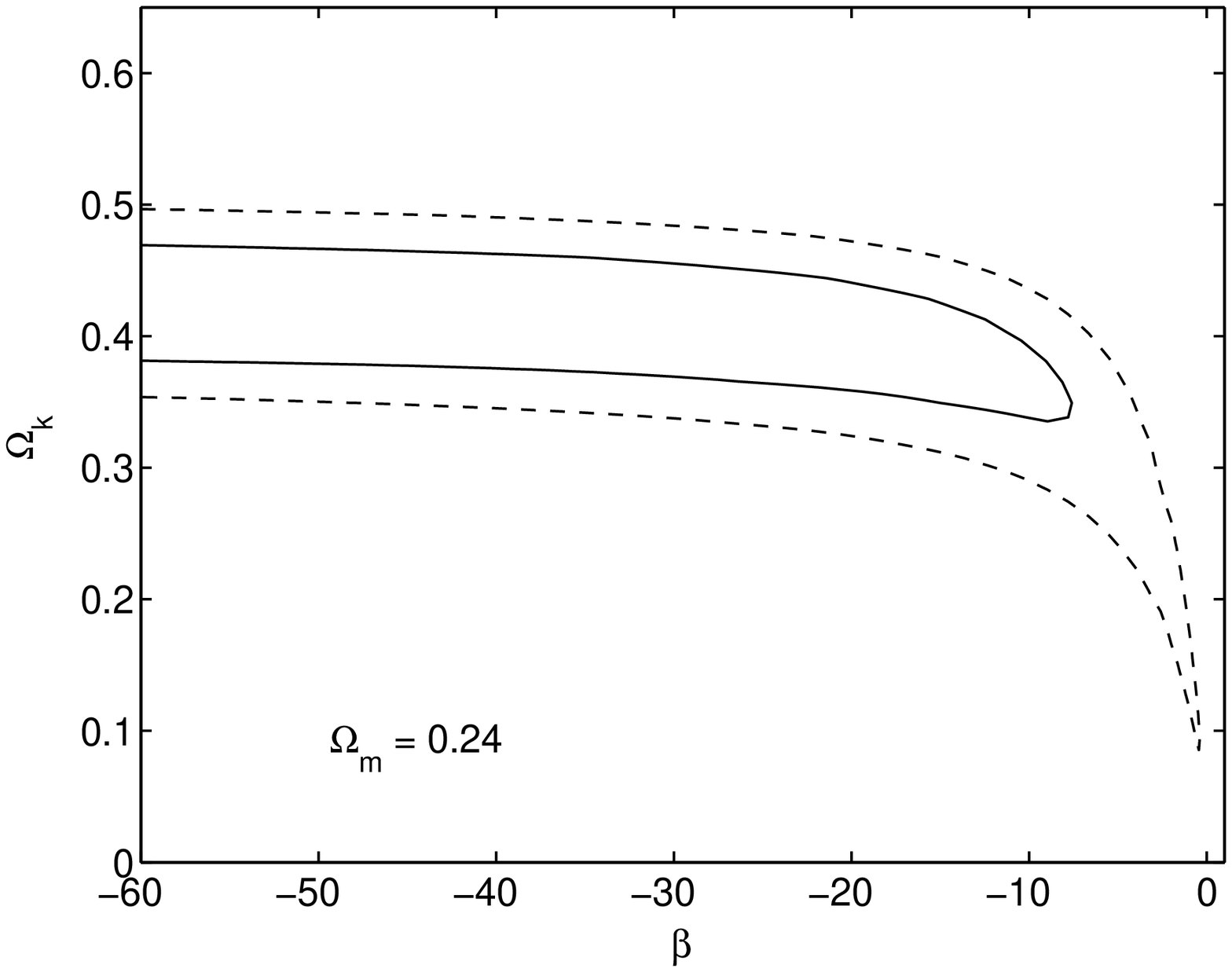}
\caption{\label{fig:DT} Confidence contours for the non-flat DT model.}
\end{minipage}
\end{figure}

If  we assume that the universe is matter dominated
 the Friedmann expansion law can  be written as
\begin{equation}
\label{FriedDTmodel} \left({ {H}\over{H_0}} \right)^2 = \Omega_{m}
(1+z)^3 + (1-\Omega_{m}-\Omega_k) \left({{H}\over{H_0}}
\right)^\beta \nonumber \\
+ \Omega_k (1+z)^2 ~.
\end{equation}
\begin{figure}[ht]
 \includegraphics[width=18pc]{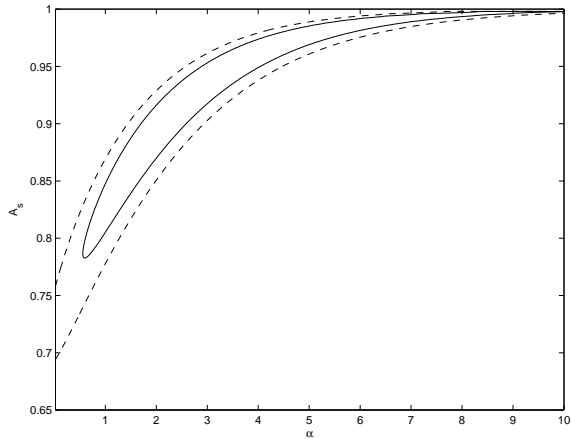}\hspace{2pc}%
\begin{minipage}[b]{16pc}  \caption{\label{fig:GCGflat} Confidence
 contours for the flat GCG model.}
\end{minipage}
\end{figure}
Notice that $\beta$ is the only parameter of the model: for
$\beta=0$ the new term behaves like a cosmological constant
and the case $\beta=1$ corresponds
to the
Dvali-Gabadadze-Porrati (DGP) model \cite{Dvali}. The
 requirement that the new term does not interfere
with the formation of large-scale structure leads to the bound $\beta \leq 1$.
\begin{figure}[h]
\begin{minipage}{37pc}
 \includegraphics[width=18pc]{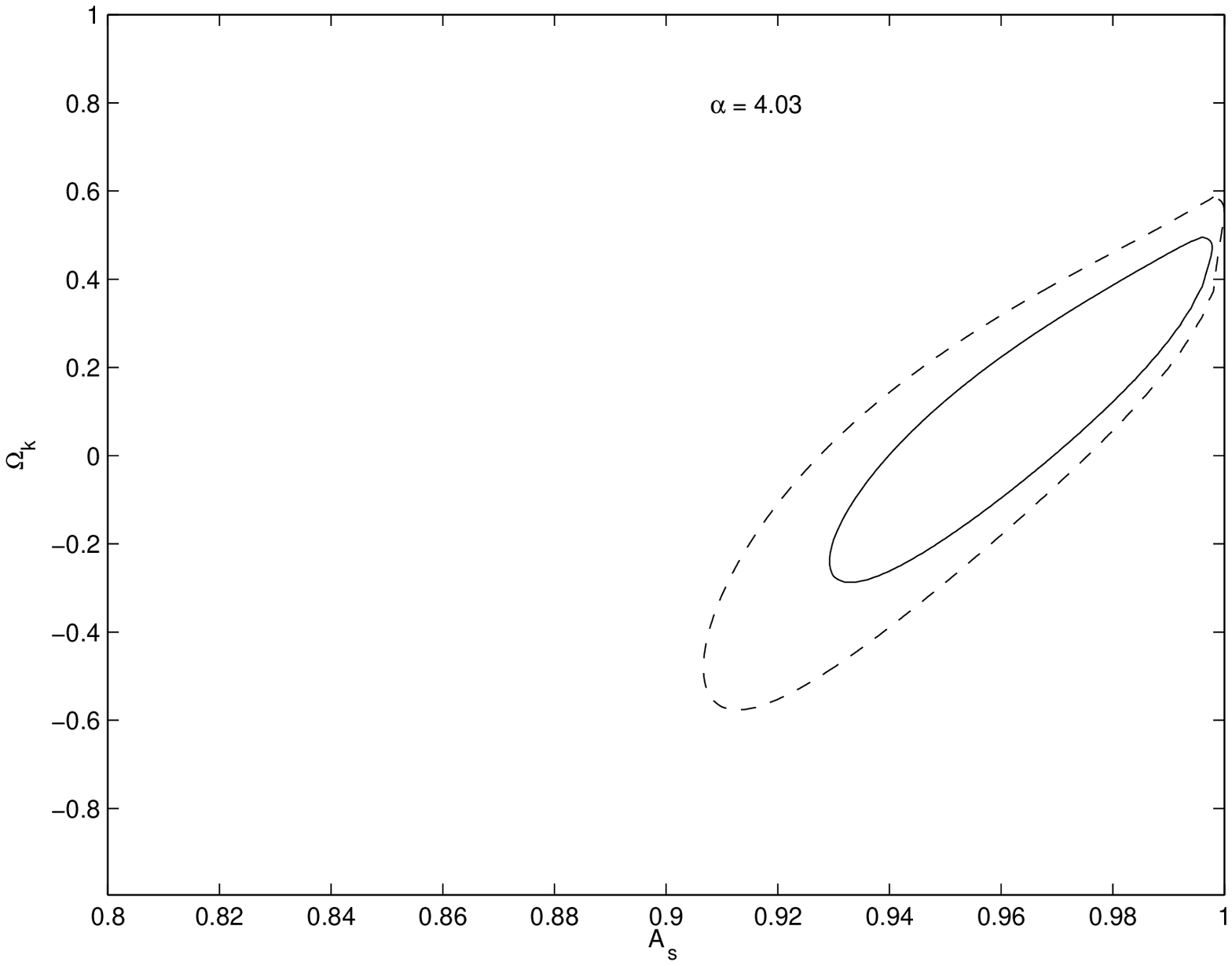}
 \includegraphics[width=18pc]{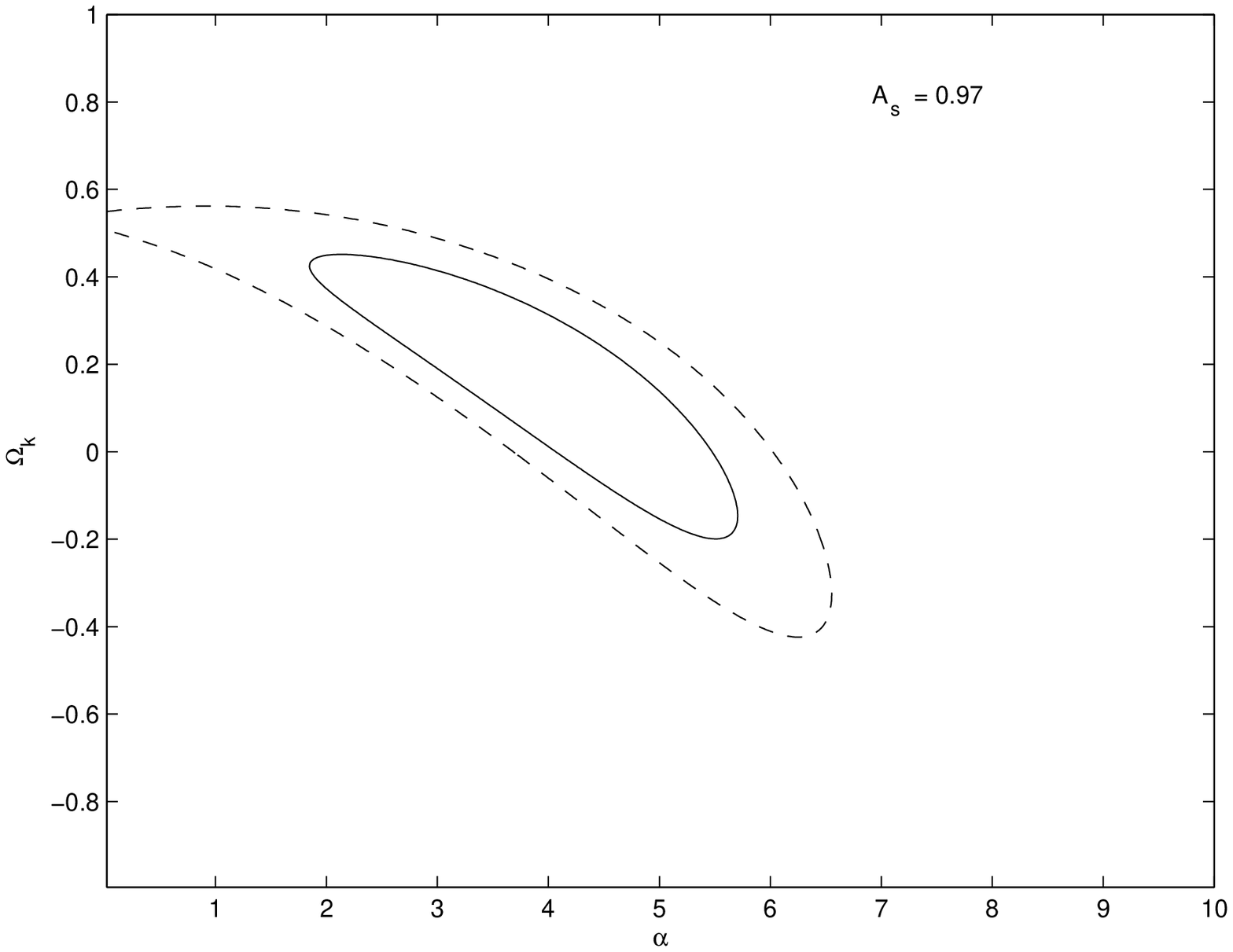}
  \caption{\label{fig:GCG} Confidence contours for the non-flat GCG model. }
\end{minipage}
\end{figure}

 From Fig. \ref{fig:flatDT},
 it is clear that $\beta$ is very weakly constrained
 and can become arbitrarily large and
negative. Moreover, both the $\Lambda$CDM model 
($\beta=0$) and  the DGP model ($\beta=1)$,
are strongly disfavoured. Fixing  $\beta=1$ (DGP model),  we get
  $\Omega_{m} = 0.17$ .

If we relax the flat prior, we find that the best fit
results are for
 $\{\Omega_{m}, \Omega_k, \beta\}= \{0.24, 0.43, -60,0\}$.
 As we can see from the contour plots shown in Fig. \ref{fig:DT},
  $\beta$ can become arbitrarily large and
negative, if we allow $\Omega_m$ to be large. Moreover, both
$\Lambda$CDM and DGP models are disfavoured at $95\%$ C.L.

\section{Generalized Chaplygin gas model}

Finally, we consider the generalized Chaplygin gas model, which is
characterized by the equation of state
\begin{equation}
p_{ch} = - {A \over \rho_{ch}^\alpha}~, \label{rhoGCG}
\end{equation}
where $A$ and $\alpha$ are positive constants. For $\alpha=1$, the
equation of state is reduced to the Chaplygin gas scenario
\cite{Kamenshchik:2001cp}.
Integrating the energy conservation equation with the equation of
state (\ref{rhoGCG}), one gets \cite{Bento:2002ps}
\begin{equation}
\rho_{ch} = \rho_{ch0} \left[A_{s} + {(1-A_s) \over
a^{3(1+\alpha)}}\right]^{1/(1+\alpha)}~,
\end{equation}
where $\rho_{ch0}$ is the present energy density of GCG and $A_s
\equiv A/\rho_{ch0}^{(1+\alpha)}$. Hence,  $\rho_{ch}$, interpolates
 between a dust dominated phase,
$\rho_{ch} \propto a^{-3}$, in the past and a de-Sitter phase,
$\rho_{ch} = -p_{ch}$, at late times. This property makes the GCG
model an interesting candidate for the unification of dark matter
and dark energy. 
 Notice  that $\alpha = 0$ corresponds to the
$\Lambda$CDM model.

The Friedmann equation for a non-flat unified GCG model
is given by
\begin{equation}
\left({ {H} \over {H_0} }\right)^2  =
\left(1-\Omega_{k}\right)\left[A_{s} +
(1-A_s)(1+  z)^{3(1+\alpha)}\right]^{1/(1+\alpha)} 
+  \Omega_{k}(1+z)^{2}~~.
\label{FriedGCGmodel}
\end{equation}

This model has been thoroughly scrutinized from the observational
point of view; for a CMB
power spectrum compatibility analysis see e.g. Ref. 
\cite{Bento2003} and for a
 previous supernovae data analysis  see Ref. \cite{Bertolami2004} and
 references therein.
 The $68\%$ and $95\%$ confidence
level contours for the flat case are  shown in Fig.
\ref{fig:GCGflat}. Notice that the $\Lambda$CDM model is consistent
 at $95\%$ C.L although it is ruled out at $68\%$ C.L.

When we relax the condition of flat prior, the best fit model 
becomes  very close to flat  ($\Omega_k = 0.02$) which is a
 a new result.  The confidence contours for
this case are shown in Fig. 6.

\section{Conclusions}

We performed a likelihood analysis of the latest
type Ia supernovae data for three  models: the modified gravity
  Cardassian and  
Dvali-Turner models and the generalized Chaplygin gas model of
unification of dark energy and dark matter. We find that SNe Ia
most recent data allows, in all cases, for non-trivial constraints
on model parameters as summarized in Table  \ref{table:best}.
  We find that, for
all models, relaxing the flatness condition implies that data
favors a positive  curvature and the GCG model is nearly flat in this
case. The fact that the  {\it gold} sample of supernovae
data prefers a flat GCG model, which is consistent with CMB
observations, leads us to conclude that  the GCG is a better choice among the
three  alternative models that we have
considered.

\end{document}